\documentclass[fleqn,twoside]{article}
\usepackage{latexsym}
\usepackage{espcrc2}
\usepackage{graphicx}
\usepackage{epsfig,rotating}
\usepackage{hyperref}

\def\NON{\nonumber\\}

\def\ie{\mbox{\it i.e.} }
\def\eg{\mbox{\it e.g.} }                      
\def\vev#1{\Big\langle #1 \Big\rangle}           

\def\svev#1{\left\langle #1\right\rangle}       
\def\leqx{\,\raisebox{-1.0ex}{$\stackrel{\textstyle <}{\sim}$}\,}

\def\bj{\overline\psi}

\def\tG{\tilde\G}

\def\gh{\hat\g_5}
\def\mres{m_{res}}
\def\tH{\tilde{H}}
\def\bar#1{\overline{#1}}

\def\eg{\mbox{\it e.g.} }

\def\svev#1{\left\langle #1\right\rangle}

\def\APH#1{Ann. Phys. {\bf #1}}

\def\NPB#1{Nucl. Phys. {\bf B#1}}
\def\NPBP#1{Nucl. Phys. B (Proc. Suppl.) {\bf #1}}
\def\PLB#1{Phys. Lett. {\bf B#1}}
\def\PRD#1{Phys. Rev. {\bf D#1}}

\def\PRL#1{Phys. Rev. Lett. {\bf #1}}
\def\PRP#1{Phys. Rep. {\bf #1}}

\let\ced=\c                     

\def\c{\chi}
\def\d{\delta}
\def\g{\gamma}

\def\j{\psi}

\def\l{\lambda}
\def\m{\mu}

\def\p{\pi}                     
\def\r{\rho}                    
\def\t{\tau}

\def\G{\Gamma}
\def\J{\Psi}

\def\ch{{\cal H}}   

\def\cl{{\cal L}}

\title{Localization in lattice QCD (with emphasis on practical implications)}

\author{
Maarten Golterman\address{
Department of Physics and Astronomy,
San Francisco State University,
San Francisco, CA 94132, USA}
and
Yigal Shamir\address{
School of Physics and Astronomy,
Raymond and Beverly Sackler Faculty of Exact Sciences,
Tel-Aviv University, Ramat~Aviv 69978, ISRAEL}%
\thanks{presenter of plenary talk at Lattice 2003, Tsukuba}
}

\begin{document}

\begin{abstract}
When Anderson localization takes place in a quenched disordered system,
a continuous symmetry can be broken spontaneously
without accompanying Goldstone bosons. Elaborating on this observation
we propose a unified, microscopic physical picture of the phase diagram of
quenched {\it and} unquenched QCD with two flavors of Wilson fermions.
The phase with Goldstone bosons---by definition the Aoki phase---is
always identified as the region where the mobility edge of
the (hermitian) Wilson operator is zero.
We then discuss the implications for domain-wall and overlap fermions.
We conclude that both formulations are valid only well outside
the Aoki phase of the associated Wilson-operator kernel,
because this is where locality and chirality can be both maintained.
\end{abstract}

\maketitle

\setcounter{footnote}{0}
\section{Introduction}
Long ago, a conjecture was made for the phase diagram of lattice QCD with
two flavors of Wilson fermions and a standard plaquette action \cite{aoki}.
(For earlier work see ref.~\cite{Jan}.)
This phase diagram is displayed in Fig.~1.
In the Aoki phase (region B in the figure), parity and isospin undergo
spontaneous symmetry breaking (SSB) by a pion condensate.\footnote{
  For $g_0>0$, chiral symmetry is explicit broken by the Wilson term,
  and $\svev{\bj\j}$ is not an order parameter.
  In the continuum limit chiral symmetry is recovered, and
  one can rotate the pion condensate back to $\svev{\bj\j}$,
  see \eg ref.~\cite{shsi}.
}
According to the Banks-Casher relation \cite{BC}
for the case at hand,
the pion condensate is the response
to an (infinitesimal) applied twisted-mass \cite{aoki,tmqcd},
\begin{equation}
  \svev{\p_3} =  2\p\r(0) \,.
\label{BC}
\end{equation}
Here $\r(\l)$ is the spectral density of the hermitian\footnote{
  Conventions: $H(m_0)= \g_5 D(m_0)$
  where the usual Wilson operator is  $D(m_0)=D^{naive} - W - m_0$.
  The sum of the bare mass $m_0$ plus the
  Wilson term $W$ is a strictly positive operator
  for $m_0>0$. The super-critical region where zero modes may exist
  is $0>am_0>-8$.
}
Wilson operator $H(m_0)$.
The twisted-mass term is $m_1\, i\bj \g_5\t_3\j$.
Inside the Aoki phase, the other two pions are Goldstone bosons
associated with SSB of isospin down to a U(1) symmetry.

Region C of the phase diagram is interesting
because, as explained below, this is where one can
use domain-wall \cite{dwf,fs} and overlap \cite{oovlp,ovlp} fermions.
According to the original conjecture \cite{aoki}, all correlation functions
of Wilson fermions are short-ranged well inside region C.
This was supported
by quenched numerical results \cite{Aokiq}.\footnote{
  {\it Exceptional} configurations \cite{excep} --
  defined by the condition that $H(m_0)$ has
  a zero mode for some $m_0$ -- were discarded in ref.~\cite{Aokiq}.
  We return to this issue below.
}
Today, we have convincing evidence that,
in region C of the {\it quenched} theory,
there is a non-zero density of small-size near-zero modes of $H(m_0)$
for any $g_0>0$ \cite{scri,bnn}.\footnote{
  Further evidence comes from the body of numerical work
  using domain-wall and/or overlap fermions.
}
This raises the following puzzle. Given that $\r(0)\ne 0$,
the Banks-Casher relation~(\ref{BC}) implies that the pion condensate is
non-zero; hence there is SSB of isospin and parity;
but since isospin is a continuous symmetry,
the Goldstone theorem requires the existence of massless Goldstone bosons.
Thus, the absence of long-range correlations appears to contradict the
existence of a density of near-zero modes in region C of the quenched theory!

\begin{figure}[thb]
\begin{center}
\epsfig{file=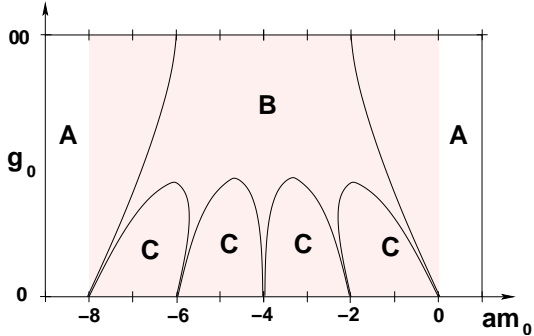, width=7.0cm}
\end{center}
\vspace*{-1.2cm}
\caption{{\it Phase diagram in the bare coupling -- bare mass plane.
The shaded area is the super-critical region.}}
\vspace*{-0.8cm}
\end{figure}

Before tackling this puzzle, it is instructive to describe the above
results using a language borrowed from
the physics of disordered systems (for reviews see ref.~\cite{cmrev}).
For this purpose we may interpret $H(m_0)$ as the hamiltonian
of a five-dimensional system, and thus its eigenvalues $\l_n$
as ``energy'' eigenvalues. The eigenfunctions $\J_n(x)$
may be thought of as the wave functions of ``electrons''
that reside on the sites $x$ of a four-dimensional spatial lattice.
{\it Disorder} is provided by the ensemble of
gauge-field configurations, in terms of which $H(m_0)$ is defined.

Fig.~2 is a cartoon aimed at explaining how the small-size,
small-$\l$ eigenstates of $H(m_0)$ emerge with increasing disorder.
An infinite volume is assumed.
The first row describes the free operator $H_0(m_0)$,
assuming for definiteness that $am_0 \approx -1$.
The gap is $O(1)$ in lattice units.
Next, consider one small dislocation (second row), by
which we mean that the link variables inside a small-size hypercube
may take any value, but all other links are still equal to one.
It was shown that a bound state -- whose eigenvalue is located
anywhere we like inside the gap -- can be produced by adjusting the links
of the dislocation \cite{bnn}. The next step is to consider a dilute
gas of small dislocations, choosing the position and shape of the individual
dislocations at random. A bound state produced by a particular dislocation
is only negligibly affected by all other dislocations.
Taken together, the (fairly) randomly distributed eigenvalues of all
the bound states fill up the gap.

The last step is to consider realistic (quenched or dynamical)
Monte-Carlo configurations.
Unlike the previous nearly-free cases,
now we cannot define an ``asymptotic value of the potential,''
nor what is the ``binding energy.''
The concept of bound states is inadequate and, instead,
we have {\it exponentially localized} states.
Likewise, we now have {\it extended} states instead of scattering states.
Still, we hypothesize that there are {\it distinct} spectral intervals
containing either extended or
localized eigenstates, but not both.\footnote{
  We are not aware of any proof of this -- widely used -- assertion.
}

Up to short-distance random fluctuations, the mode density of
an exponentially localized eigenstate behaves like
\vspace{-.5ex}
\begin{equation}
  | \J_n(x) |^2 \sim
    {\exp\left(- |x-x_n^0| / l_n \right)\over l_n^4} \,.
\label{lclx}
\end{equation}
\vspace{-.5ex}
On average, the mode density decays exponentially away from
some ``center'' $x_n^0$, where
$l_n$ is the {\it localization length}.\footnote{
  In four dimensions, typically $\J_n(x)$ will be localized inside a region
  whose volume is $O(l_n^4)$. This explains the factor of $1/l_n^4$
  on the right-hand side of eq.~(\ref{lclx}), for a normalized eigenstate.
}
As we increase $|\l_n|$ starting from $\l_n \sim 0$,
tunneling becomes easier and, on average, $l_n$ increases.
The localization length diverges when $|\l_n|$
reaches the {\it mobility edge} $\l_c$,
and for $|\l_n|>\l_c$ the eigenstates become extended.\footnote{
  Since $H_0(m_0)$ is bounded, we expect another
  mobility edge $\bar\l_c$, such that,
  for $|\l_n|>\bar\l_c$, the eigenstates
  are again localized (lightly shaded area to the very right in Fig.~2).
}

\begin{figure}[bth]
\vspace*{-0.5cm}
\begin{center}
\begin{turn}{-90}
\epsfig{file=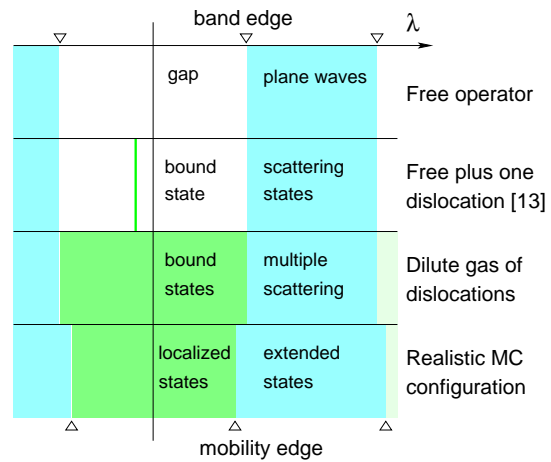, width=6.0cm}
\end{turn}
\end{center}
\vspace*{-1.0cm}
\caption{{\it Spectrum of $H(m_0)$.
On the horizontal axis are the eigenvalues $\l_n$.
The gross features of the spectrum are symmetric around $\l=0$.
Part of the negative-$\l$ spectrum is not shown.}}
\end{figure}
%

The value of the mobility edge is a dynamical feature.
For $g_0\ll 1$ it will be close to the band edge of $H_0(m_0)$,
but for larger $g_0$ it may go down and, eventually, even vanish
(for $am_0\sim -1$).

\section{Localization, mobility edge and the phase diagram}

Let us now reconsider the above puzzle. Assume first that the density
of near-zero modes of $H(m_0)$ arises from extended states only ($\l_c=0$).
Since these eigenstates spread throughout the entire lattice,
some long-range correlations may be expected,
and the pion condensate should be accompanied by Goldstone bosons.
In contrast, assume that the condensate arises
from exponentially localized eigenstates only  ($\l_c>0$).
These eigenstates are sensitive only to the link variables
inside a very small lattice volume (centered around $x_n^0$),
and long-range correlations are unlikely.
We are now closer to resolving the puzzle: if $\r(0)$ arises from
exponentially localized eigenstates only, indeed we {\it should} not
expect any long-range correlations.

What about the Goldstone theorem?
The relevant Ward identity is
\begin{equation}
  \sum_\m (1+O(ap))\, p_\m\tG_\m(p) + 2m_1 \tG(p)
  = \svev{\p_3} \,,
\label{WI}
\end{equation}
where $\tG(p)$ and $\tG_\m(p)$ are the Fourier transforms of
$\svev{\p_+(x)\,\p_-(y)}$ and $\svev{J_\m^+(x)\,\p_-(y)}$ respectively.
$J_\m^+(x)$ is a conserved isospin current.
(The $\pm$ refer to operators that raise/lower isospin
by one unit.) As we explain below,
when SSB takes place in a {\it quenched} theory,
it is possible to saturate the Ward identity without Goldstone bosons.
Instead, the two-point function of the would-be Goldstone bosons {\it diverges}
like $1/m_1$ in the limit $m_1\to 0$ \cite{mcks,paper}.
In the unquenched theory, on the other hand, the product $m_1 \tG(p)$
vanishes in this limit for any $p\ne 0$,
and the Goldstone theorem applies.

In the rest of this section we expand on the main physical issues.
More results, as well as missing technical details,
may be found in ref.~\cite{paper}, henceforth denoted $I$.

After a change of variables $\bj'=\bj \g_5$ (see also footnote~2),
the fermion lagrangian reads $\cl = \bj' (H(m_0)-im_1\t_3)\j$.
Let us first consider the finite-volume quenched theory,
and look for a possible $1/m_1$ divergence of the
(position space) pion two-point function $\G(x,y)$.
In terms of the spectrum of $H(m_0)$ we have
\begin{equation}
  \G(x,y)
  = \svev{\sum_{kn}
  {\J_n^\dagger(x) \J_k(x)\; \J_k^\dagger(y) \J_n(y)
  \over (\l_k + im_1)(\l_n - im_1)}
  } \,.
\label{dblsum}
\end{equation}
The terms labeled by $k$ ($n$) refer to the up (down) quark.
$\svev{\cdot}$ denotes the functional integration over the gauge field
with the Boltzmann weight of the quenched theory.
In the denominators, the $im_1$ terms provide the prescription
for integrating around the poles at $\l_k=0$ or $\l_n=0$ for $m_1\to 0$.
The only terms which may diverge in this limit are those where $\l_k=\l_n$,
\ie those where $k=n$ (ignoring accidental degeneracies).
Keeping these terms only we have, for $m_1\to 0$,
\begin{eqnarray}
  \G(x,y)
  &\!\! \approx \!\! & {1\over m_1}
  \svev{ \sum_n |\J_n(x)|^2 |\J_n(y)|^2 {m_1 \over \l_n^2 + m_1^2}}
\NON
  &\!\! \leadsto \!\! &{\p \over m_1}
  \svev{ \sum_n |\J_n(x)|^2 |\J_n(y)|^2\; \d(\l_n)}
\nonumber
\end{eqnarray}
In a (large but) finite volume, $H(m_0)$ has near-zero modes practically
everywhere inside the super-critical region
(see $I$ for a more detailed discussion).
Since the quenched-theory Boltzmann weight is strictly positive and
independent of $m_1$, the pion two-point function
indeed has a $1/m_1$ divergence!
Clearly, (nearly) exceptional configurations \cite{excep} -- the ones
responsible for the near-zero modes of $H(m_0)$ --
are also directly responsible for the $1/m_1$ divergence.
We conclude that the finite-volume, super-critical quenched theory
is ill-defined for $m_1=0$, so one must keep $m_1 \ne 0$.

By summing the previous equation over $x$ and then using translation
invariance to average over $y$ we find, for $m_1\to 0$,
\begin{equation}
  m_1 \, a^4 \! \sum_x \G(x,y)
  \to {\p \over V} \bigg\langle\sum_n\d(\l_n)\bigg\rangle = \p \r(0) \,,
\label{Grho}
\end{equation}
which verifies the Ward identity~(\ref{WI}) for $p=0$ in finite volume.
The $1/m_1$ divergence allows for $\r(0) > 0$ and, hence, SSB,
in the {\it finite-volume} quenched theory.\footnote{
  In the unquenched case one can prove that $\G(x,y)$ is bounded.
  The $m_1\to 0$ limit of eq.~(\ref{Grho}) then yields zero, which is the familiar
  result that there is no SSB in finite volume.
}

While conceptually important, the previous discussion provides little
insight into the origin of the $1/m_1$ divergence.
In fact, we believe that in the {\it infinite-volume} limit
this divergence arises only from exponentially localized eigenstates.
We will illustrate the relevant physics via a heuristic argument:
provided $\r(0)$ arises from localized states
only, there are no Goldstone poles. To this end,
consider the Fourier transform of the mode density
$\ch_n(p) = a^4\! \sum_x |\J_n(x)|^2 \, e^{-ipx}$. In view of eq.~(\ref{lclx})
we assume the {\it ansatz}
\begin{equation}
  \ch_n(p) \approx {e^{-ipx_n^0}\over 1 + p^2 l_n^2} \,.
\label{Hn}
\end{equation}
The essential point is that, for $p^2 l_n^2 \ll 1$, one should feel
only the exponentially decaying envelope of the mode density
({\it cf.}\ eq.~(\ref{lclx})) but not the short-distance fluctuations.
The phase factor in the numerator reflects the location of the mode.
The normalization implies $\ch_n(0)=1$.
(Positivity of the mode density implies $|\ch_n(p)| \le 1$,
so $p=0$ is the (global) maximum of $|\ch_n(p)|$.
This forbids a linear term in $p$ in the denominator of eq.~(\ref{Hn}).)
For $m_1 \to 0$ and $p^2 l_n^2 \ll 1$ we now obtain
\begin{eqnarray}
  2m_1\tG(p) &=& {2\over V}
  \svev{\sum_n |\ch_n(p)|^2\;   {m_1 \over \l_n^2 + m_1^2}}
\NON
  & \leadsto & 
  2\p\r(0) \left(1 + O\! \left( p^2\, \overline{l}^2 \right)\right) \,,
\label{noGB}
\end{eqnarray}
where $\overline{l}$ is the (suitably defined) average localization length
of the near-zero modes. Using eq.~(\ref{WI}) we conclude that
$\tG_\m(p) \sim ip_\m \r(0)\, O(\overline{l}^2)$, implying that
there are no Goldstone poles.

In conclusion, we arrive at the following physical conjecture
for $m_1 \to 0$ in the super-critical region
as the simplest one that fits all the existing evidence.
We define the Aoki phase as the region
where pseudo-scalar Goldstone bosons exist.
This region coincides with the part of the phase diagram
where $\l_c=0$. Inside (and close to the boundaries of) the Aoki phase,
the chiral lagrangian should provide a valid description
of the long-range physics \cite{shsi,gss}.
Outside the Aoki phase $\l_c>0$, and
the spectrum of localized eigenstates extends down to $\l=0$.
So far, this picture applies to
{\it both} the quenched and the unquenched theories.
The difference is that, for $\l_c>0$, the quenched theory
has $\r(0)>0$ and a divergent pion two-point function;
the unquenched theory has $\r(\l) \sim \l^2$ for small $\l$
due to the fermion determinant,
and a condensate can only be built from extended eigenstates
(\ie when $\l_c=0$).

\section{Domain-wall and overlap fermions}

Domain-wall fermions (DWF) and overlap fermions are
sophisticated descendants of Wilson fermions, that allow separation
of the chiral and the continuum limits. They are recognized today
as closely-related realizations of the Ginsparg-Wilson (GW) relation
\cite{GW}. They are both built around
a Wilson-operator ``kernel'' $H(-M)$,
where $0<aM<2$ is the domain-wall height.\footnote{
  Here we assume a fifth-dimension lattice spacing $a_5 \le 1$.
}
DWF are five-dimensional Wilson fermions,
in which only hopping terms in the four physical directions couple
to four-dimensional link variables, which themselves are independent of the
fifth coordinate \cite{dwf,fs}. The latter takes values $s=1,2,\ldots,N_s$.
The left- and right-handed components of the quark field
are localized on opposite boundaries of the five-dimensional space,
and chiral symmetry becomes exact in the limit $N_s\to\infty$ \cite{fs}.

Let $T(M,a_5)$ be the transfer matrix that hops the fermions
in the fifth dimension. The ``hamiltonian'' $\tH(M,a_5)=-\log(T(M,a_5))/a_5$
is closely related to the Wilson operator.
One has $\tH(M,0)=H(-M)$. Also, the spectrum of exact zero modes
is independent of $a_5$. We may thus study the approach
to the chiral limit using the physical concepts of
the previous section. We begin by considering
the DWF's PCAC relation (the superscript $a$ is an SU(N)-flavor index)
\cite{fs}
\begin{equation}
   \sum_\m \partial^*_\m A^a_\m(x) =
   2m_q J^a_5(x) + 2 J^a_{5q}(x)\,.
\label{pcac}
\end{equation}
Here $A^a_\m(x)$ is the partially-conserved DWF axial current,
$\partial^*_\m$ is the backward derivative, and
$m_q$ is the bare quark mass. The pseudo-scalar density
$J^a_5(x)$ is localized on the boundaries of the fifth dimension
and serves as an interpolating field for pions. $J^a_{5q}(x)$ is another
pseudo-scalar density located in the middle of the fifth dimension,
which gives rise to finite-$N_s$ chiral symmetry violations.
A measure of these violations is the residual mass
$\mres=\mres(\t,N_s)$, which may be defined as
\begin{equation}
  \mres =
  {\sum_{\vec{x}\vec{y}\t'}\vev{J^+_{5q}(\vec{x},\t+\t')\,J^-_5(\vec{y},\t')}
            \over
  \sum_{\vec{x}\vec{y}\t'}\vev{J^+_5(\vec{x},\t+\t')\, J^-_5(\vec{y},\t') }}\,.
\label{mres}
\end{equation}
Here we singled out one lattice direction as euclidean time $\t$.
For large $\t>0$ and large $N_s$, we expect,
ignoring power corrections,\footnote{
  In a numerical simulation, the signal of localized modes becomes more
  complicated when full translation invariance is not enforced.
}
\begin{equation}
  \mres \,\sim\, c_1 \exp(-\t(1/\bar{l} - m_\p)) + c_2\, q^{N_s} \,.
\label{mresx}
\end{equation}
This should be true provided $\l'_c>0$,
where $\l'_c=-\log(q)/a_5$ is the mobility edge of $\tH(M,a_5)$;
equivalently, $0<q<1$. Let us briefly explain
this result (for  the limitations of our analysis, see $I$).
The denominator in eq.~(\ref{mres}) is governed by a zero-momentum pion
and decays like $\exp(-m_\p\t)$. In the numerator,
one should distinguish between the contributions of
extended and of localized modes of $\tH(M,a_5)$.
For large $N_s$, modes with eigenvalue near $\l'_c$ dominate
the extended modes' contribution.
This results in universal fifth-dimension wave functions
for the left- and right-handed quarks,
which decay exponentially with $s$ \cite{next,paper}.
The extended modes' contribution
thus factorizes like $q^{N_s} \exp(-m_\p\t)$.
The contribution of localized modes to the numerator
is not suppressed exponentially with $N_s$, because the localized
spectrum goes down to zero. This contribution can be estimated
using the asymptotic form~(\ref{lclx}). It will be dominated by
localized near-zero modes centered around the straight line connecting
the two space-time points (see $I$ for details) and is estimated to
behave like $\exp(- \t /\bar{l})$ where now $\bar{l}$ is,
in effect, the maximal localization length of the near-zero modes.

According to eq.~(\ref{mresx}), $\mres$ does not vanish exponentially with $N_s$,
for any {\it fixed} value of $\t$. However,
let us imagine taking the infinite-volume limit while keeping the lattice
spacing fixed. If we extract $\mres$ using increasingly large $\t$,
the first term on the right-hand side of eq.~(\ref{mresx}) will ultimately vanish
even for fixed $N_s$, provided $1/\bar{l}>m_\p$.
Therefore, the localized modes' contribution
to $\mres$ does not necessarily constitute
a violation of chiral symmetry.

The overlap operator \cite{oovlp,ovlp} corresponds to the double limit
$N_s\to\infty$ and $a_5\to 0$ of DWF. Explicitly,
\begin{equation}
  aD_{ov}= 1 - \g_5 \gh\,,\ \ \
  \gh \equiv {H(m_0)\over |H(m_0)|}\,.
\label{Dovlp}
\end{equation}
This operator satisfies the GW relation \cite{GW},
\ie it possesses a modified chiral symmetry
(with the same algebraic properties as ordinary chiral symmetry).
The overlap operator cannot have a finite range \cite{horvath},
and the relevant question is what are its localization properties.
Exponential locality of the overlap operator
was proved in ref.~\cite{hjl},
provided all the plaquette variables are uniformly bounded
close to one (``admissibility condition'').
In this case the spectrum of $H(m_0)$ has a gap,
as on the first row of Fig.~2.\footnote{
  Ref.~\cite{hjl} generalized the proof to the case depicted on
  the second row of Fig.~2.
}
In realistic MC simulations, however,
one cannot impose such a constraint on the plaquettes.
Assuming the mobility edge satisfies $\l_c>0$,
the spectrum looks like the last row of Fig.~2.
The effect of the localized spectrum can be analyzed along similar lines.
Considering the restriction of $\gh$ to (say) localized modes
with $|\l|\le \l_c/2$, denoted $\gh^<$, we estimate
\begin{equation}
  \Big\langle|\gh^<(x,y)|\Big\rangle
  \leqx
  \exp(-|x-y|/(2 \bar{l})) \,,
\label{projRR}
\end{equation}
where now $\bar{l}$ is determined by all eigenmodes with $|\l|\le \l_c/2$.
Once the near-zero modes do not hamper locality,
exponential locality of $D_{ov}$
can be established as in ref.~\cite{hjl}.
Finally, a similar analysis may be applied to
the $N_s\to\infty$ limit of DWF with $a_5>0$, which
is also described by eq.~(\ref{Dovlp}), except that $H(m_0)$ is replaced by
$\tH(M,a_5)$ \cite{limNs} (see also ref.~\cite{AB}).

The main conclusion of the previous discussion is the following:
DWF and overlap fermions can be used only {\it well outside}
the Aoki phase of their Wilson kernel.
This situation is quenched-like:
(even) for (dynamical) DWF or overlap simulations,
the Boltzmann weight does not contain
the determinant of the Wilson operator itself.
Nevertheless,  the spectral properties of the Wilson-operator kernel
are crucial, and, for {\it any} ensemble of gauge-field configurations,
we may (formally) introduce two quenched Wilson flavors
and look for their Aoki phase.

When the mobility edge of the Wilson kernel
is $O(1)$ in lattice units, chiral symmetry of DWF will be recovered
exponentially with increasing $N_s$, for large $\t$.\footnote{
  The situation is more complicated when considering the lattice
  renormalization of the effective Electro-Weak hamiltonian.
  Especially when power-divergent subtractions are involved,
  localized modes may still give rise to chiral symmetry violations
  which are {\it not} suppressed by any space-time separation.
  In this case, employing smeared links \cite{fat} and/or
  the ``projection method'' of ref.~\cite{kjkn}
  may provide better control over the subtractions
  needed to recover chiral symmetry.
}
Well outside the Aoki phase we also expect the average
localization length of the near-zero modes to be
$O(1)$ in lattice units, and this guarantees the locality of the overlap
operator, as well as of the generalized overlap operators obtained
from DWF with $a_5>0$. In contrast, being inside the Aoki phase
means that the mobility edges of both the Wilson kernel and $\tH(M,a_5)$
are zero. This corresponds to taking the limits $q \to 1$ and
$\bar{l}\to\infty$, where eqs.~(\ref{mresx}) and~(\ref{projRR}) cease to hold.
Thus, DWF will develop long-range correlations in all five dimensions,
and the (generalized) overlap operator
defined by the $N_s\to\infty$ limit will become non-local,
including for $a_5 \to 0$.

Near the continuum limit, DWF and overlap fermions are local.
The mobility edge and $1/\bar{l}$ both scale like the lattice cutoff,
and $\r(0)$ tends to zero rapidly for $g_0\to 0$ \cite{scri}.
But, for present-day simulations, it is unclear if $1/\bar{l}$
is large compared to \eg the rho-meson mass.
How, then, can one tell if a given set of simulation parameters lies
safely outside the Aoki phase?
At the very least, one should require
that $1/\bar{l}$ will be bigger than the mass of all hadrons
of interest.  The issue definitely deserves further study.

For DWF, eq.~(\ref{mresx}) shows that, by monitoring the dependence of $\mres$
on both $N_s$ and $\t$, one can extract the crucial spectral properties
of $\tH(M,a_5)$. The mobility edge is determined in terms of $q$,
and may be extracted from the $N_s$ dependence.
The (dominant) localization length of the near-zero modes
may be extracted from the $\t$ dependence, and the constant $c_1$
provides information on the density of the near-zero modes.
A well-behaved $\mres$ ({\it cf.}\ eq.~(\ref{mresx}))
also ensures the locality of the effective four-dimensional
operator defined by the $N_s\to\infty$ limit.
$\mres$ is routinely calculated in any new DWF simulation.
For a lattice cutoff $a^{-1} \sim 2$ GeV,
well-behaved $\mres$ have been obtained in quenched simulations
with Iwasaki \cite{CPPACS} and DBW2 \cite{RBC} gauge actions.
Likewise, when overlap fermions are employed,
it should become a routine practice to determine the localization properties
of the overlap operator! In particular, the relation between the
localization lengths of the overlap and of the
near-zero modes should be studied in more detail.

Dynamical DWF \cite{dyn} and overlap simulations are, and will be,
very expensive. In a dynamical simulation the danger is
that one will end up too close to, or even inside, the Aoki phase.
In this case, one {\it cannot}
maintain both chirality and locality.\footnote{
  We believe that this applies to present-day thermodynamical simulations
  carried out with a lattice cutoff $a^{-1} \sim 1$ GeV.
}
Having to give up on something, we believe that one should insist on locality,
at the price of doing worse on chirality.
The reason is that, unlike with approximate chiral symmetry
(see below), very little is known about how
to monitor for the physical consequences of deteriorating locality.
Any non-locality of an operator that satisfies the GW relation
should be regarded as the source of an unknown systematic error.

In a situation where the overlap operator
(or generalizations to $a_5>0$) is non-local,
locality could be maintained, for example, by using
DWF with modest $N_s$.\footnote{
  If no pseudo-fermion fields are included, this is strictly local.}
For other approximate solutions of the GW relation see ref.~\cite{GWx}.\footnote{
  Local approximations of the overlap exist, too.
  For example, a rational polynomial approximation may be represented
  as a continued fraction which, in turn, may be cast in the form
  of a five-dimensional action with nearest-neighbor coupling \cite{cf}.
  This action is {\it not} yet local because it contains parameters
  which are functions of the minimal
  eigenvalue of $H(m_0)$, which in turn is a {\it global}
  property of the gauge-field configuration. If one fixes these parameters,
  an ultra-local five-dimensional action is obtained.
  }
While violations of chiral symmetry in Ward identities will be non-negligible
in this case, the well-established renormalization program allows us
to subtract them. In principle, this is the same situation as with
ordinary Wilson fermions. In practice, however, unlike Wilson fermions,
as long as $\mres$ is a few MeVs (or less),
it may be possible to carry out
the subtractions successfully for a wide range of weak matrix elements.

\vspace{2ex}\noindent
{\it Acknowledgements}. We thank the organizers for a well-organized,
pleasant meeting, and
Philippe de Forcrand for discussions of the overlap operator.
YS is supported by the Israel Science Foundation under grant
222/02-1. MG is supported in part by the US Department of Energy.

\end{document}